\documentclass[nofootinbib,amsfonts,prd,aps]{revtex4}
\usepackage{graphicx}
\begin{document}

\title{Spherically symmetric gravity coupled
  to a scalar field with a local Hamiltonian: the complete
  initial-boundary value problem using metric variables}

\author{Rodolfo Gambini$^{1}$,
Jorge Pullin$^{2}$}
\affiliation {
1. Instituto de F\'{\i}sica, Facultad de Ciencias, 
Igu\'a 4225, esq. Mataojo, Montevideo
, Uruguay. \\
2. Department of Physics and Astronomy, Louisiana State University,
Baton Rouge, LA 70803-4001}

\begin{abstract}
  We discuss a gauge fixing of gravity coupled to a scalar field in
  spherical symmetry such that the Hamiltonian is an integral over
  space of a local density. In a previous paper we had presented it
  using Ashtekar's new variables. Here we study it in metric
  variables. We specify completely the initial-boundary value problem
  for ingoing Gaussian pulses.
\end{abstract}

\maketitle

\section{Introduction}
Spherically symmetric gravity coupled to a scalar field is a rich
model, where one can test scenarios of black hole formation, the
critical phenomena discovered by Choptuik and Hawking evaporation at
the quantum level. For many years the full quantization of the model
resisted analysis, in part due to the complicated nature of the
Hamiltonian structure of the system.  Initial attempts to study the
problem were done by Berger, Chitre, Nutku and Moncrief \cite{bcnm}
and further developed by Unruh \cite{unruh}. The resulting complicated
nature of the gauge fixed Hamiltonian led led Unruh to say ``I present
it here in the hope that someone else may be able to do something with
it.'' More recently, Husain and Winkler and Daghigh, Kunstatter and
Gegenberg \cite{huwi}, using Painlev\'e--Gullstrand coordinates
simplified somewhat Unruh's treatment. None of these efforts provided
a Hamiltonian that was the spatial integral of a local density,
leading to non-local equations of motion with the ensuing difficulty
at the time of quantization.

We recently noted that using Ashtekar's new variables the construction
of a local Hamiltonian was possible. It was later suggested by Unruh
\cite{unruhpersonal} and Gegenberg and Kunstatter \cite{geku} that a
similar construction was possible in metric variables. In hindsight,
this is not too surprising. The key element used in our construction
was that in Ashtekar's variables the gravitational part of the
Hamiltonian constraint becomes the total derivative of a quantity with
respect to the radial variable. It turns out that some years ago,
Kucha\v{r} \cite{kuchar} introduced canonical coordinates for
spherically symmetric vacuum gravity in which one of the coordinates
is the mass as function of the radius. The gravitational part of the
Hamiltonian constraint in that case is given by the total derivative
of the mass with respect to the radial variable. Therefore a
construction similar to the one we had carried out with Ashtekar's
variables can be carried out with Kucha\v{r}'s variables. We will
detail the construction here. As most gauge fixings, only certain
families of initial data can be accommodated with a given choice of
gauge. We set up a suitable initial-boundary value problem in the
gauge fixed theory and for the physically important case of Gaussian
pulses.

The organization of this paper is as follows. In section II we discuss
the gauge fixing in terms of Kucha\v{r}'s variables. In section III we
set up the Hamiltonian.  In section IV we study the
initial-boundary value problem, in particular for Gaussian pulses. We
end with a discussion of possibilities for quantization.

\section{Gauge fixing in the Kucha\v{r} variables}

The starting point is the three-metric in spherical coordinates,
\begin{equation}
  ds^2 = \Lambda(r)^2 dr^2+R(r)^2 
\left(d\theta^2 +\sin^2\theta d\varphi^2\right),
\end{equation}
with $\Lambda$ and $R$ arbitrary functions of the radial variable (and
time), and their corresponding canonical momenta $P_\Lambda$ and
$P_R$. The canonical formulation in terms of these variables has been
discussed by Kucha\v{r} \cite{kuchar}, so we will not repeat it here,
we refer the reader to his paper for details. The total Hamiltonian
density is obtained from the Hamiltonian and diffeomorphism
constraints,
\begin{eqnarray}
  H_T &=& N H + N^r C_r,\\
  H &=& \frac{1}{G}\left( -\frac{\Lambda}{2} -\frac{P_\Lambda P_R}{R}+
\frac{P_\Lambda^2 \Lambda}{2 R^2}+\frac{\left(R'\right)^2}{2\Lambda}
-\frac{R' R \Lambda'}{\Lambda^2}+\frac{R"
  R}{\Lambda}\right)+\frac{P_\phi^2}{2 R^2\Lambda}+\frac{R^2 
\left(\phi'\right)^2}{2\Lambda},\\
  C_r &=&\frac{1}{G}\left(-P_\Lambda' \Lambda+P_R R'\right)+P_\phi \phi',
\end{eqnarray}
with $N$ the lapse and $N^r$ the shift. 

We now proceed to redefine the lapse and shift
\begin{eqnarray}
  N_{\rm old} &=& \frac{N_{\rm new} R'}{\Lambda},\\
  N^r_{\rm old} &=& N^r_{\rm new} +\frac{N_{\rm old} P_\Lambda}{R'R},
\end{eqnarray}
and from now on we drop the ``new'' subscripts. The total Hamiltonian
density can then be written with the gravitational part explicitly as a
derivative with respect to the radial coordinate,
\begin{equation}
  H_T = N\left\{ \frac{1}{G}\left( \frac{\left(R'\right)^2 R}{2 \Lambda^2}
      -\frac{R}{2} -\frac{P_\Lambda^2}{2 R}\right)'
+\frac{R'}{2\Lambda^2R^2}\left[ P_\phi^2 + R^4
  \left(\phi'\right)^2\right]
+\frac{P_\Lambda P_\phi \phi'}{\Lambda R}\right\}+ N^r
\left[\frac{1}{G}\left(-P_\Lambda'\Lambda+P_R R'\right)+P_\phi \phi'\right].
\end{equation}

Let us now proceed to gauge fix. We start by setting $R=r$.
Preserving this condition in time implies that the shift vanishes
(this is the rescaled shift, the original shift does not vanish). One
solves the diffeomorphism constraint to obtain $P_R$. 

To completely fix the gauge, we need to fix another variable. With
that objective in mind, it is convenient to rewrite the Hamiltonian
as,
\begin{equation}
  H_T=-N X' +N G\left(\frac{P_\phi^2}{2 \Lambda^2 r^2} +\frac{r^2
      \left(\phi'\right)^2}{2 \Lambda^2} + 
\frac{P_\Lambda P_\phi \phi'}{r\Lambda}\right),\label{8}
\end{equation}
with
\begin{equation}\label{9}
  X = -\frac{r}{2\Lambda^2} +\frac{r}{2} +\frac{P_\Lambda^2}{2r} - \frac{R_S}{2},
\end{equation}
with $R_S$ at the moment just a constant, later it will be identified
with the Schwarzschild radius. 

The strategy for finding a gauge fixing that leads to a local
Hamiltonian will be to fix the value of the quantity $X$. The
resulting constraint therefore depends on the gravitational variables
undifferentiated. When one preserves that gauge fixing in time, the
lapse will be fixed by an algebraic equation rather than a
differential one. This is the key point. If one were left as usual
with a differential equation, the lapse would be an integral of the
canonical variables. Since the Hamiltonian is an integral that
involves the lapse,  it
becomes an integral of an integral and in that sense is non-local. So we
choose $X=f(r,t)$. Preservation in time of this condition determines
the lapse as an algebraic function of $\phi,P_\phi,\Lambda,P_\Lambda$.

We proceed to solve the variable $\Lambda$ through the gauge fixing, 
\begin{equation}
  \Lambda=
 - \frac{r}{Y},
\end{equation}
with
\begin{equation}
  Y=\sqrt{r^2+P_\Lambda^2-r R_S -2
    f(r,t) r}
\end{equation}
and substitute it in the total Hamiltonian, which leads to,
\begin{equation}\label{12}
 f'=\frac{G\left(P_\phi^2+\left(\phi'\right)^2
     r^4\right)\left(r^2+P_\Lambda^2-r R_S-2 f(r,t) r\right)}{2 r^4}-
\frac{G  P_\Lambda P_\phi \phi' Y}{r^2}
\end{equation}
which we should solve to get $P_\Lambda$ as a function of $\phi$ and
$P_\phi$. We will see later how to do this in a compact way.

This completes the gauge fixing. The free variables are
$\phi,P_\phi$. We now go to the evolution equations for those
variables, derived before the gauge fixing, and substitute the
latter. The resulting equations can be shown to be equivalent to those
that stem from a true Hamiltonian,
\begin{equation}
  H_{\rm True} = \dot{f}(r,t)\frac{r^2}{G Y \left(P_\Lambda(\phi,P_\phi)+Y\right) },
\end{equation}
with 
\begin{equation}
  Y=\sqrt{r^2+P_\Lambda^2(\phi,P_\phi)-r R_S -2
    f(r,t) r}
\end{equation}
and in these expressions $P_\Lambda$ should be substituted by the
expressions we derived before during the gauge fixing. 

\section{Obtaining the true Hamiltonian directly}

A constructive procedure to directly obtain the above true Hamiltonian
is to perform a canonical transformation from the variables
$\Lambda,P_\Lambda$ to a new set of variables $X,P_X$. This should be
done before the gauge fixing $X=f(r,t)$, so at the moment $X$ is  function of
$\Lambda$ and $P_\Lambda$ given by (\ref{9}). 
This motivates us to consider a generating function of
type 3, $F_3(P_\Lambda,X)$ for which one would have that,
\begin{equation}
  \Lambda = \frac{\partial F_3(P_\Lambda,X)}{\partial P_\Lambda},
\end{equation}
and solving for $\Lambda$ in the definition of $X$ (\ref{9}) 
this can be integrated to give 
\begin{equation}
  F_3(X,P_\Lambda) = -\frac{r}{G} \ln\left(P_\Lambda+Z\right),
\end{equation}
with $Z$ being the non-gauge fixed version of $Y$,
\begin{equation}
  Z=\sqrt{r^2+P_\Lambda^2(\phi,P_\phi)-r R_S -2 X r}.
\end{equation}
We therefore have for the conjugate variable,
\begin{equation}
  P_X = -\frac{\partial F_3(P_\Lambda,X)}{\partial X}= 
\frac{r^2}{G Z\left(P_\Lambda+Z\right)},
\end{equation}
and for $\Lambda$, 
\begin{equation}
    \Lambda = \frac{\partial F_3(P_\Lambda,X)}{\partial P_\Lambda}=
 - \frac{r}{\sqrt{r^2+P_\Lambda^2 -r R_S -2 X r}}, \label{17}
\end{equation}
The total Hamiltonian in terms of the new variables is
\begin{equation}
  H_{\rm tot} =   N\left(P_X + \frac{r^2}{G Z\left(P_\Lambda+Z\right)}\right). 
\end{equation}
with $P_\Lambda$ obtained by solving (\ref{8}) with $\Lambda$
substituted by (\ref{17}). 

We now proceed to gauge fix $X=f(r,t)$. Preservation in time of this
condition leads to $N=\dot{f}(r,t)$. Noting that $\phi,P_\phi$ have
vanishing Poisson brackets with $P_X$, if we write the evolution
equations and substitute the gauge fixing in them, we have that,
\begin{eqnarray}
  \dot{\phi}&=&\left\{\phi,\int dr H_{\rm
      tot}\right\}=\left\{\phi,\int dr H_{\rm true}\right\},\\
  \dot{\phi}&=&\left\{\phi,\int dr H_{\rm
      tot}\right\}=\left\{\phi,\int dr H_{\rm true}\right\},
\end{eqnarray}
with 
\begin{equation}
  H_{\rm true} = \dot{f}(r,t) \frac{r^2}{G Y\left(P_\Lambda(\phi,P_\phi)+Y\right)}.
\end{equation}

\section{Setting initial and boundary data}

As in any gauge fixing in a complicated theory like general relativity,
one does not expect one will cover all of phase space. The limitation
here is given by the equation for $P_\Lambda$, which written explicitly reads,
\begin{equation}
a P_\Lambda^4+b P_\Lambda^2+c =0,\label{24}
\end{equation}
with 
\begin{eqnarray}
  a&=&\frac{G^2}{4r^4} \left({P_\phi^2} -{r^4}\left(\phi'\right)^2\right)^2,
\\
  b&=&G W\left({P_\phi^2} +{r^4}\left(\phi'\right)^2\right)
-{G^2 P_\phi^2 \left(\phi'\right)^2\left(r^2 -r R_S -2 f(r,t)
    r\right)},\\
  c&=&r^4 W^2,\\
  W&=& -f'(r,t)
  +\frac{G\left(P_\phi^2+r^4\left(\phi'\right)\right)
\left(r^2 -r R_S -2 f(r,t)\right)}{2r^4}.\label{28}
\end{eqnarray}
This will not generically yield a real
value for $P_\Lambda$ given arbitrary initial data for
$\phi,P_\phi$. This not only limits the initial data but also the
boundary conditions one can give at the outer and inner boundary. So
from now on we are limited to consider more specific situations. One
has certain freedom to modify things by playing with the function
$f(t,r)$ that determines the gauge fixing. For different choices of
$f(t,r)$ different families of initial and boundary data will be
acceptable as producing real values for the variables.

A case of great interest is the study of the propagation of wave
packets of scalar field on a black hole space-time. We
will therefore concentrate ourselves on that situation. This will
require specifying at spatial infinity boundary conditions such that the
geometry is asymptotically that of Schwarzschild with no ingoing
matter fields, and the inner boundary corresponds to a dynamical
horizon with matter fields purely ingoing into it. 

We choose as initial data for the scalar field,
\begin{equation}
  \phi(t=0,r)= \phi_0 \frac{\exp\left(-\frac{\left(r-r_0\right)^2
        \sigma^2}{4}\right) \sigma\sqrt{\pi} }{r} 
\frac{\left(r-R_S\right)^2}{r^2},
\end{equation}
where we are considering a Gaussian pulse and we added a factor
$(r-R_s)^2/r^2$ such that the field vanishes at the horizon
initially. This makes the horizon for the initial data an isolated
one. For $P_\phi$ we choose what is needed to have a purely ingoing
pulse,
\begin{equation}
  P_\phi(t=0,r)= 
   \phi_0 \frac{\exp\left(-\frac{\left(r-r_0\right)^2
        \sigma^2}{4}\right) \sigma\sqrt{\pi} }{2r}
  \frac{\left(r-R_S\right)^2}{r^2}\frac{\left(\sigma^2 r^2 +2 -\sigma^2 r r_0\right)}{r^2}.
\end{equation}

We will now proceed to fix the gauge in such a way that the
bi-quadratic equation (\ref{24}) has at least a pair of real
roots. Notice that in (\ref{24}) both $a$ and $c$ are always
positive. For the quadratic equation for $P_\Lambda^2$ have a positive
root one needs to make the linear term negative. One possible strategy
is to consider (\ref{28}) and integrate it using the initial data we
are considering, 
\begin{equation}
  f'(r,t)=   \frac{G\left(P_\phi^2+r^4\left(\phi'\right)\right)
\left(r^2 -r R_S -2 f(r,t)\right)}{2r^4}.\label{28}  
\end{equation}
The right hand side is a bit complicated, but evaluating it
numerically one sees it has the form of a Gaussian-like shape. One can
therefore simply take for $f$ a Gaussian that envelops the integral as
a gauge choice,
\begin{equation}
  f' = G \sigma^5
\left(r-R_S\right)^3 \exp\left(-\frac{\left(r-r_0\right)^2
        \sigma^2}{2}\right),\label{32}
\end{equation}
with $k$ appropriately big for it to envelop the integrand. The
integral for $f$ can be evaluated in closed form, but its expression
is lengthy. The form of the function is relatively simple, it is a
modified step function as shown in the figure.
\begin{figure}[ht]
\begin{center}
\includegraphics[height=6cm]{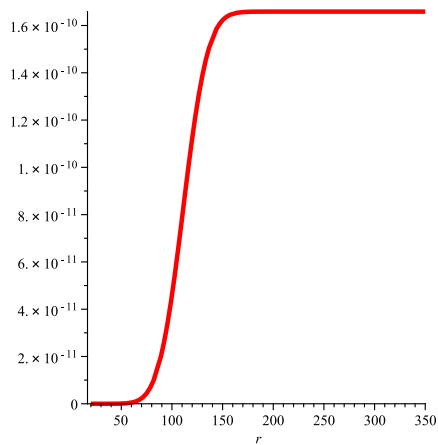}
\end{center}
 \caption{The function providing the gauge fixing, shown for
   $\sigma=0.05$, $G=10^{-11}$, $R_S=2$, $r_0=100$.}
\end{figure}
In terms of $f$ one can solve (\ref{24}) for $P_\Lambda$. The closed
form expression is again lengthy. Asymptotically for large $r$ we have
$P_\Lambda=c_0 r^{3/2}$ with $c_0$ a constant. 

This completes the determination of the initial data. We need to fix
the gauge for all time. To do this, we consider the preservation of
$f(t,r)=X$, which determines the lapse. We would like the lapse, at
least asymptotically, to reproduce the usual manifestly asymptotically
flat nature of the Schwarzschild space-time. This corresponds to $N=1$
(recall that we are referring to $N_{\rm new}=1$, which corresponds
asymptotically to $N_{\rm old}=1/\Lambda$). This results,
asymptotically in $X=X_{\rm asymp.}+A(r) t$, where $X_{\rm asymp.}$ is
the asymptotic value of the expression of $f$, which, for instance,
can be read off for large values of $r$ in the figure above. The
expression of $A(r)$ is,
\begin{equation}
  A(r)= c_1 r^4 \exp\left(-\frac{1}{2}\left(r-r_0\right)^2\sigma^2\right),
\end{equation}
with $c_1$ a constant. 

With this form of the gauge fixing, the asymptotic form of the metric
is,
\begin{eqnarray}
  g_{tt}&=&-1+\frac{2GM}{r},\\
  g_{tr}&=&\frac{P_\Lambda}{\sqrt{r^2+P_\Lambda^2-2GMr}},\\
    g_{rr}&=&\frac{r^2}{r^2+P_\Lambda^2-2GMr},
\end{eqnarray}
with $GM=R_S/2+X_{\rm asymp.}$, which shows that the asymptotic mass
is the same as that of the horizon plus the contribution of the scalar
field. With a simple redefinition of $t$ this yields the usual
expression of the Schwarzschild metric in the Schwarzschild
coordinates. 

Although we have not studied the evolution in detail, one can envision
using a gauge with (\ref{32}) modified to be an ingoing pulse and this
should yield real expressions for all quantities as the pulse travels
inward, at least far away from the black hole.

\section{Discussion}

We have shown that one can gauge fix spherically symmetric gravity
coupled to a scalar field in terms of the traditional metric variables
with a Hamiltonian that is the integral of a local density in explicit
form. We construct a family of gauge fixings that can accommodate
ingoing Gaussian pulses and show that they include manifestly
asymptotically flat coordinates. The construction of the gauge is such
that it is clear that it will evolve correctly in time, at least for a
limited amount of time. It should be noted that we have not analyzed
properly the inner boundary condition beyond the initial slice. One
presumably would like to have a dynamical horizon that increases its
mass as the ingoing pulses progress towards the black hole, at least
studying the problem classically. Quantum mechanically, it is less
clear what one needs at the horizon, since Hawking radiation should be
present. 

The question of quantization of the model in terms of this gauge
implies having to deal with the quartic equation (\ref{24}) that
generically will lead to complex values. It is therefore unclear that
the constructed Hamiltonian could be promoted to a self-adjoint
operator. It should be noted that there exist techniques
\cite{techniques} to deal with these types of issues in quantization.
We have recently illustrated this in a model system
\cite{nosotrosmodelo}.  They are however, limited to certain regimes.
Realistically, this type of approach is unlikely to yield insights
about extreme regimes like the ones close to the singularity. But it
may be useful in other situations, like in those in which a large
black hole emits Hawking radiation to study, for instance, the back
reaction of the weak radiation on the large black hole.

\acknowledgements

This work was supported in part by grant NSF-PHY-0968871, funds of the
Hearne Institute for Theoretical Physics, CCT-LSU, Pedeciba and ANII
PDT63/076. This publication was made possible through the support of a
grant from the John Templeton Foundation. The opinions expressed in
this publication are those of the author(s) and do not necessarily
reflect the views of the John Templeton Foundation.

\end{document}